\documentclass{article}

\usepackage{arxiv}

\usepackage[utf8]{inputenc} % allow utf-8 input
\usepackage[T1]{fontenc}    % use 8-bit T1 fonts
\usepackage{hyperref}       % hyperlinks
\usepackage{url}            % simple URL typesetting
\usepackage{booktabs}       % professional-quality tables
\usepackage{amsfonts}       % blackboard math symbols
\usepackage{nicefrac}       % compact symbols for 1/2, etc.
\usepackage{microtype}      % microtypography
\usepackage{lipsum}		% Can be removed after putting your text content
\usepackage{graphicx}
\usepackage[square,numbers]{natbib}
\usepackage{doi}
\usepackage{adjustbox}
\usepackage{xcolor}
\usepackage{tcolorbox}
 
\usepackage{enumerate}
\usepackage{multirow}
\usepackage{amssymb}
\usepackage{multirow}
\usepackage{graphicx}
\usepackage{amsmath}
\usepackage{cases}
\usepackage{array}
\usepackage{url}
\usepackage{booktabs}

\title{TranSalNet: Towards perceptually relevant visual saliency prediction}

\date{} 					% Or removing it

\author{ 
Jianxun Lou \\
School of Computer Science and Informatics\\
Cardiff University\\
United Kingdom, CF24 4AX \\
\\
\And
Hanhe Lin \\
National Subsea Centre\\
Robert Gordon University\\
United Kingdom, AB10 7AQ \\
\\
\And
David Marshall \\
School of Computer Science and Informatics\\
Cardiff University\\
United Kingdom, CF24 4AX \\
\\
\And
Dietmar Saupe \\
Department of Computer and Information Science\\
University of Konstanz\\
Germany, 78464 \\
\\
\And
Hantao Liu \\
School of Computer Science and Informatics\\
Cardiff University\\
United Kingdom, CF24 4AX \\
\\
}

\begin{document}
\maketitle

\begin{abstract}
Visual saliency prediction using transformers~\footnote{This paper has been published in \textit{Neurocomputing}, Volume 494, 14 July 2022, Pages 455-467, DOI: \url{https://doi.org/10.1016/j.neucom.2022.04.080} under the CC BY 4.0 license (Open Access).} - Convolutional neural networks (CNNs) have significantly advanced computational modelling for saliency prediction. However, accurately simulating the mechanisms of visual attention in the human cortex remains an academic challenge. It is critical to integrate properties of human vision into the design of CNN architectures, leading to perceptually more relevant saliency prediction. Due to the inherent inductive biases of CNN architectures, there is a lack of sufficient long-range contextual encoding capacity. This hinders CNN-based saliency models from capturing properties that emulate viewing behaviour of humans. Transformers have shown great potential in encoding long-range information by leveraging the self-attention mechanism. In this paper, we propose a novel saliency model that integrates transformer components to CNNs to capture the long-range contextual visual information. Experimental results show that the transformers provide added value to saliency prediction, enhancing its perceptual relevance in the performance. Our proposed saliency model using transformers has achieved superior results on public benchmarks and competitions for saliency prediction models.

The source code of our proposed saliency model TranSalNet is available at: \url{https://github.com/LJOVO/TranSalNet}.
\end{abstract}

% keywords can be removed

\section{Introduction}
Visual attention represents an important mechanism of the human visual system (HVS), which allows humans to select and interpret the most relevant information in the visual scene \cite{eyefix_Jonides}. Simulating visual attention in the form of an algorithm is regarded as visual saliency prediction. Being able to automatically predict saliency is beneficial to many research fields including computer vision, robotics, healthcare, and multimedia~\cite{Borji_2018, SONG2018218, HAN2021705, 6963384, CHEN202159, HAN201370, MISHRA202195}.

\begin{figure*}
\centering
\includegraphics[width=0.9\textwidth]{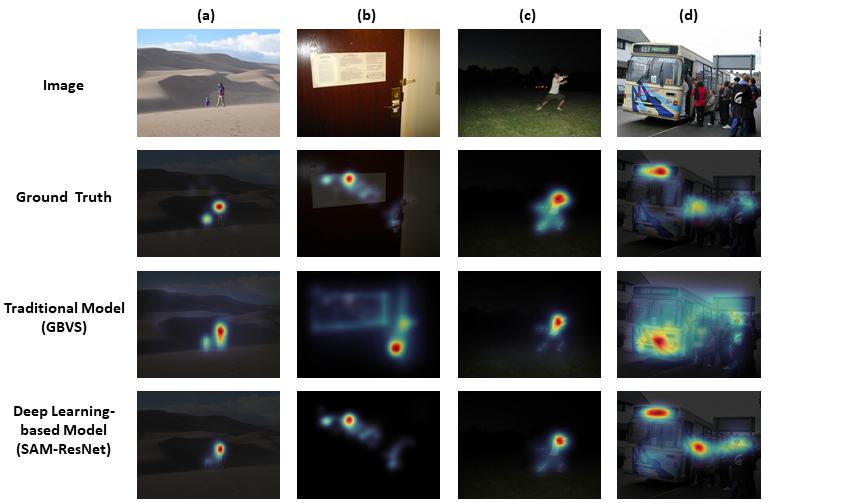}
\caption{Examples of visual saliency prediction. The first row shows the images that stimulate the human eye to view freely. The so-called ``Ground Truth" in the second line refers to the fixation density maps, also called saliency maps, generated from the human fixation location. The third and fourth rows show the prediction results of the traditional (GBVS) and deep learning-based (SAM-ResNet) saliency models, respectively. Image (a) and (b) are from MIT1003 dataset; (c) and (d) are from SALICON dataset. It can be seen that both traditional and deep learning-based models are capable of capturing human viewing behaviour, but the deep learning-based model provides better results in demanding scenes, such as (b) and (d) to a considerable extent.}
\label{fig:demo_intro}
\end{figure*}

Existing saliency prediction models can be categorized into two types, traditional and deep learning-based models. Traditional models~\cite{ITTI_2006, GBVS_2006, CovSal_2013, LDS} apply low-level visual features such as colour, luminance, texture, and contrast, to simulate the visually salient areas in the scene. These models remain rather limited as higher-level features such as objects are often omitted; but these features exhibit significant determinants of visual saliency \cite{STOLL201536,Wolfgang_2008}. Although some traditional models \cite{traditional_p1} have been extended with specific higher-level features, e.g., faces and texts, there are still obstacles in combining low-level and higher-level visual features. Rather than designing handcrafted features, deep learning-based saliency models automatically discover representations from images \cite{ML-Net, DVA_Wang, SAM_Cornia, MSI-Net, EML-NET, UNISAL, CASNET2, DeepGaze2, DP2E}. These methods typically use convolutional neural networks (CNNs) to construct feature encoders and decoders to generate visual saliency maps. Deep learning-based visual saliency models have achieved remarkable success, mainly due to the availability of well-established deep CNNs \cite{AlexNet, VGG, RESNET, DenseNet} and large-scale datasets relevant to human visual attention \cite{salicon2015}. Figure~\ref{fig:demo_intro} illustrates examples of visual saliency prediction using both traditional and deep learning-based models, and the correspondences between the ground truth (i.e., where humans look in an image) and prediction (i.e., output of a computational saliency model).

Accurately predicting saliency as perceived by humans remains an academic challenge. One way to improve the reliability of saliency prediction is to incorporate the properties of the HVS in the construction of computational models~\cite{ITTI_2006, Toet_2011}. Despite the significant progress made by the deep learning-based models, each convolution kernel in CNNs only receives information from a local subset of pixels in an image, which makes fully CNN-based models deficient in obtaining long-range contextual information. When humans view an image, foveal vision provides the highest resolution visual information, but in the meantime peripheral vision still provides the HVS with non-detailed but \textcolor{black}{long-range} visual information \cite{peripheral_1,peripheral_2,longrange_vis}. In other words, the HVS uses the \textcolor{black}{long-range} information of an image to modulate the local maxima of saliency in the visual field~\cite{Toet_2011, ITTI20001489}. Therefore, the ground truth saliency map represents the perceptual spatial interactions of local and \textcolor{black}{non-local (i.e., long-range)} information. This HVS property could be beneficial for predicting visual saliency in a perceptually more relevant manner so that the machine generated saliency map can faithfully reflect human perception. Previous studies mainly attempted to solve this problem through two approaches. One approach is capturing multi-scale information through the CNNs \cite{Huang_2015, CASNET2, MSI-Net, GAZEGAN, DVA_Wang}, which introduces image representations with different granularities to some extent. This approach may not provide the optimal solution as it still lacks the ability to model the way visual information is perceived by the HVS, e.g.,
\textcolor{black}{some studies have used multi-scale images or multi-scale representations to improve saliency prediction ~\cite{Huang_2015, CASNET2, MSI-Net, GAZEGAN, DVA_Wang}, but challenges remains for models in optimally fusing multi-scale information to mimic the functionality of the HVS.}
Another approach is adding long-range modelling capabilities to network structures to increase spatial representations. By using a Long-Short Term Memory (LSTM)-based architecture \cite{SAM_Cornia, DSCLSTM}, this approach has proven effective in handling local and \textcolor{black}{long-range} visual information thus refining the accuracy of saliency prediction. Although these studies have demonstrated promising outcomes, much work is needed to close the gap between saliency prediction and human perception. The transformer~\cite{att_is_all}, which consists of a self-attention mechanism, provides an elegant solution to process long-range information. By effectively modelling long-range dependency, the transformer has proven efficacy in the field of natural language processing~\cite{BERT}
and more recently achieved promising results in computer vision tasks~\cite{ViT,iGPT}. However, the use of transformers in visual saliency prediction has not been fully explored until now.

\textcolor{black}{To address the above-mentioned challenges and to build a human-like saliency model, we propose a novel saliency prediction model called TranSalNet, which integrates transformers into a CNN-based architecture. Transformer encoders can learn spatially long-range dependencies by using a self-attentive mechanism, resulting in a perceptually more relevant saliency representation. To the best of our knowledge, this is the first study to explore the combination of CNNs and transformers to enhance saliency prediction. Also, we demonstrate the benefits of transformer components in saliency prediction. Our model achieves superior performance not only on the MIT300 benchmark (the most widely recognised dataset for saliency benchmark) but also on the SALICON Saliency Prediction Challenge (the largest dataset available for saliency prediction).}

\section{Related work}

We contribute towards a perceptually more relevant saliency prediction method using deep learning models with transformers.
This section provides a comprehensive review on deep learning-based saliency prediction models , methods for evaluating saliency models (especially evaluating the perceptual relevance of saliency prediction), transformer applications in vision tasks, 
and multi-scale and long-range information in visual saliency prediction.

\subsection{Deep learning-based visual saliency prediction}

\textcolor{black}{A number of deep learning-based visual saliency prediction models have been proposed in recent years.
The \textit{Ensembles of Deep Networks} (eDN)~\cite{eDN} represents one of the first models that adopts shallow CNNs to detect the visual saliency of natural images. The saliency features are extracted by CNNs and combined by a linear classifier to create saliency maps. Since then, with the application of deep neural networks and large-scale saliency datasets, deep learning-based saliency prediction has achieved further remarkable successes.
\textit{DeepGaze} and \textit{DeepGaze II}~\cite{DeepGaze2}, which are based on AlexNet~\cite{AlexNet} and VGGNet~\cite{VGG}, respectively, successfully build pre-trained networks as feature extractors to train deeper networks for saliency prediction. 
By comparing VGGNet, AlexNet, and GoogleNet \cite{GoogleNet}, Huang et al.~\cite{Huang_2015} found that VGGNet detects saliency more effectively than the other two models.
Many visual saliency prediction models based on VGGNet have since been proposed~\cite{ML-Net,DVA_Wang,MSI-Net}.
\textit{EML-NET}~\cite{EML-NET} focuses on exploring the use of more sophisticated feature extractors (i.e., a parallel two-stream CNN-based encoder) to enhance the performance of saliency prediction. By comparing ResNet-50 with DenseNet~\cite{DenseNet} and NASNet~\cite{NASNet}, it is argued that in the field of saliency prediction, the widely used ResNet-50 could still be ``shallow" for the large-scale saliency datasets, such as SALICON. Similarly, \textit{DeepGaze II-E} discusses the contribution of different backbones to saliency prediction. It is found that appropriately concatenating multiple backbones pre-trained on ImageNet~\cite{imagenet} is effective in improving the performance of saliency models.}

\textcolor{black}{
In addition to the efforts mentioned above, there are several studies that adopt multi-scale or long-range information to improve visual saliency prediction. We discuss this issue below in Section ~\ref{sec:multiscale}.
}

\subsection{Evaluation methods for saliency models}    
A number of metrics have been proposed to measure the agreement between the predicted saliency map and the ground truth produced by human eye movements. By investigating commonly used metrics, Bylinskii et.~al.~\cite{Bylinskii_2019} found that under general assumptions the \textit{Linear Correlation Coefficient} (CC) and \textit{Normalized Scanpath Saliency} (NSS) metrics could be used as representative metrics for benchmarking saliency models. More importantly, they also suggested different evaluation metrics should be used for different applications, for example, metrics that are more appropriate for evaluating the capability of salient object detection may not be necessarily useful for the evaluation of saliency prediction of other vision applications ~\cite{Bylinskii_2019,xiaohan}. Li et.~al.~\cite{LIJ} found that only a limited number of evaluation metrics, i.e., NSS, CC, and \textit{Similarity} (SIM) are in close agreement with human judgements through a large-scale subjective experiment. Similarly, Yang et.~al.~\cite{xiaohan} found that CC and SIM are the most in line with human evaluation of saliency maps. Kummerer et.~al.~\cite{kummerer2018} also demonstrated that it is difficult for a saliency model to perform equally well on all popular saliency evaluation metrics. They proposed a novel approach that allows a saliency model to generate different ``saliency maps" according to the characteristics and behaviours of different metrics; and the model that adopts this evaluation method is referred to as a ``probabilistic model." As a distinction, without targeting any specific evaluation metric, a saliency model that generates a single saliency map for a given image is referred to as a ``classical model". Since our aim is to generate a single saliency map for each image that can faithfully reflect human perception, we evaluate models in the ``classical model" framework. In evaluating models, we apply all commonly used evaluation metrics to quantify model performance, but make a clear distinction between ``perception-based metrics (i.e., NSS, CC, and SIM)" and ``non-perception-based metrics", as defined in ~\cite{Bylinskii_2019}. By doing so, the perceptual relevance of the predicted saliency maps can be appropriately measured. 

\subsection{Transformer in visual tasks}

The transformer was first introduced to the tasks of natural language processing (NLP)~\cite{att_is_all}. Because of its powerful long-range dependency modelling capabilities,  the transformer has achieved remarkable success in the field of NLP. Consequently, a number of studies in the field of computer vision are also exploring the effectiveness of the use of transformer. 

The \textit{vision transformer} is one of the first pure transformer architectures for image processing, which uses a vanilla version of the transformer to form a network that achieves performance comparable to that of state-of-the-art CNN-based models. After this work, several models, such as \textit{DeepViT}~\cite{DeepViT} and \textit{Swin Transformer}~\cite{Swin_Transformer}, have achieved further success in visual tasks by using the transformer.

Currently, the transformer has also demonstrated excellent performance in the field of salient object detection~\cite{Visual_Saliency_Transformer}, which is related to the current work, even though it is a substantially different task~\cite{TPAMI_PRVS}. Salient object detection aims to segment salient objects from an image and generate a binary map~\cite{TIP_SOD}. However, in visual saliency prediction, the aim is to predict the density map of human fixations (i.e., the spatial deployment of visual attention).

In summary, the previous studies have shown the powerful representation capabilities of the transformer, particularly for capturing long-range information, which could have potential contributions to predicting gaze. However, the use of transformers in visual saliency prediction has not been fully explored until now. In this paper, we will investigate the benefits as well as application of transformer components in saliency prediction.

\subsection{Multi-scale and long-range information in visual saliency prediction}
\label{sec:multiscale}
\textcolor{black}{
By using multi-scale image representations to simulate different perceptual scales, successful results have been achieved in vision tasks such as image segmentation~\cite{unet}, human pose estimation~\cite{HRN}, and salient object detection~\cite{8782147}. In the filed of visual saliency prediction, Huang et al.~\cite{Huang_2015} and Fan et al.~\cite{CASNET2} proposed CNN-based models that extract multi-scale features from images of different resolutions separately and concatenate the results to obtain salient semantic objects with different granularities hence to optimise saliency prediction.
In order to obtain multi-scale contextual information, \textit{Deep Visual Attention} (DVA)~\cite{DVA_Wang} constructs three decoders of different granularities to generate multi-scale saliency estimates for saliency prediction.
\textit{EML-NET}~\cite{EML-NET} also uses multi-scale feature maps from encoder networks to obtain holistic scene features for saliency prediction.
\textit{MSI-Net}~\cite{MSI-Net} adopts convolutional layers with different dilation rates to augment multi-scale information for saliency prediction.
\textit{GazeGAN}~\cite{GAZEGAN} is a generative adversarial network for saliency prediction, which uses a modified U-Net with multi-scale information by using skip-connections to construct its generator.
\textit{UNISAL}~\cite{UNISAL} adopts skip-connections to provide the decoder network with multi-scale features. These studies have demonstrated that multi-scale information is beneficial to visual saliency prediction.} 

\textcolor{black}{
Similarly to other vision tasks~\cite{Visual_Saliency_Transformer,Rthnk_ss_cvpr2021}, visual saliency prediction has also benefitted from neural networks with long-range modelling capabilities to simulate the spatial attentional mechanisms.
\textit{DSCLSTM}~\cite{DSCLSTM} extracts local feature maps by using CNNs first, and then incorporates non-local scene contexts into the local feature maps by using LSTM-based components to predict human eye fixation points in natural scenes.
Cornia et al.~\cite{SAM_Cornia} developed visual saliency models that integrate an LSTM module into the CNN-based network to simulate explicit properties of the human attention mechanism. 
Similarly, Fang et al.~\cite{FANG_LSTM} used LSTM to obtain pseudo sequential information to simulate the human visual attention shift.
These studies suggest that modelling the relevant dependence between spatial information can refine the saliency prediction models.}

\textcolor{black}{
In this paper, we combine these two strategies. 
More specifically, we integrate transformer encoders into a CNN-based architecture to provide multi-scale image representations with enhanced long-range contextual information, 
resulting in perceptually more relevant visual saliency prediction.}

\section{The proposed model}

\begin{figure*}
\centering
\includegraphics[width=1.0\textwidth]{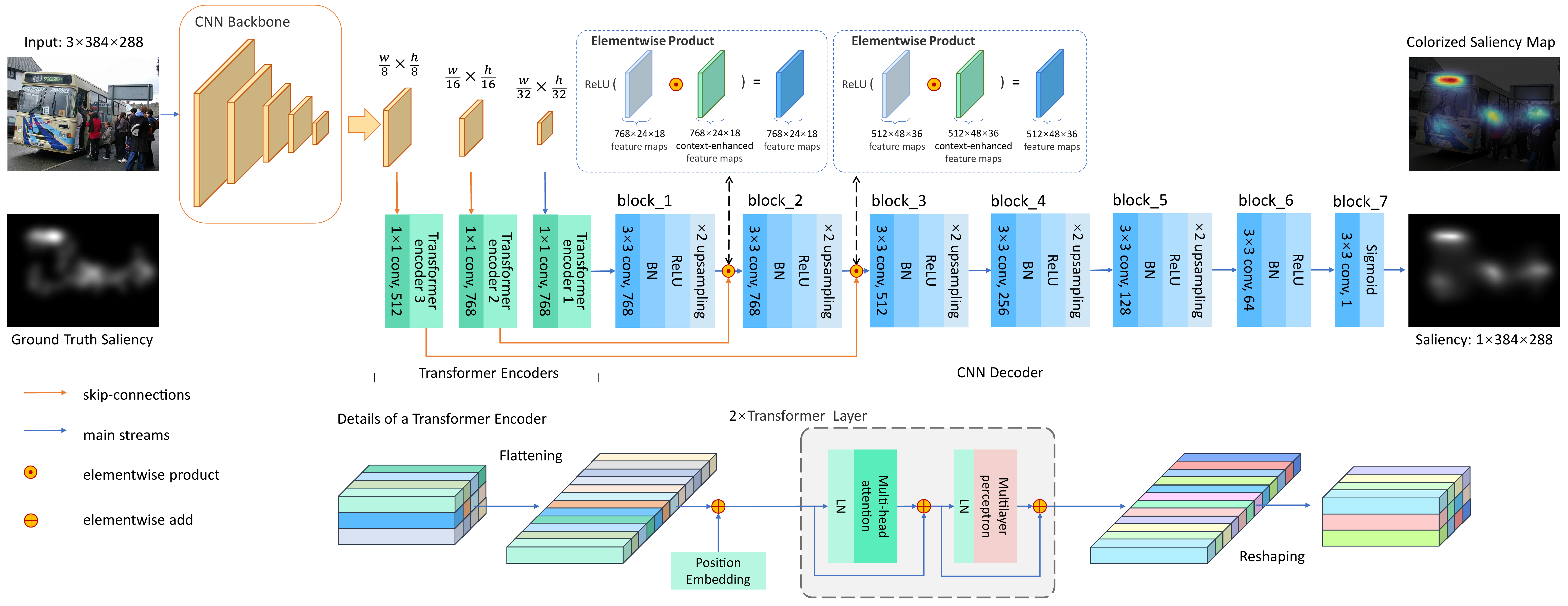}
\caption{Schematic overview of TranSalNet. Assume that the spatial size of inputs is $w \times h$. After the input image is processed by the CNN encoder, which provides three sets of multi-scale feature maps have spatial size of $\frac{w}{8}\times\frac{h}{8}$, $\frac{w}{16}\times\frac{h}{16}$, and $\frac{w}{32}\times\frac{h}{32}$, respectively. Then the contextual information of these feature maps is enhanced by transformer encoders. The predicted saliency map is generated by the CNN decoder, which uses skip-connection (\textcolor{black}{orange arrows}) and element-wise production to fuse multi-scale context-enhance feature maps. The illustration of the transformer encoder is shown below the architecture diagram, which consists of standard Multi-head Self-Attention (MSA) and Multi-layer Perceptron (MLP) blocks.}
\label{fig:arch}
\end{figure*}

The schematic overview of our proposed TranSalNet model is shown in Figure~\ref{fig:arch}. 
Firstly, a given image is fed into a CNN encoder. In order to obtain multi-scale image representations, three sets of feature maps with different spatial sizes are extracted from the CNN encoder. Due to the inherent inductive biases of CNN encoder architectures, the extracted image representations lack long-range contextual information, which potentially makes a saliency model less humanlike (note the human visual system is proficient in capturing both local and \textcolor{black}{long-range} visual information). Therefore, to obtain perceptually more relevant visual saliency prediction, these feature maps are passed through three transformer encoders, yielding long-range context-enhanced feature maps. Then the CNN decoder fuses these feature maps for saliency prediction.  

\subsection{The CNN encoder}
Previous research has shown that the use of CNN-based networks to extract features for saliency prediction is effective. Likewise, we used a CNN encoder as the feature extractor in this study.

The CNN models used in this study were initially constructed for image classification. In order to provide image feature maps to the downstream networks, the fully connected layer at the end of these CNNs is removed to form a viable CNN encoder. We extract feature maps with three sets of different spatial sizes from the CNN encoder. Given an input image with size $w \times h \times 3$, the spatial dimensions of the extracted feature maps are $\frac{w}{8}\times\frac{h}{8}$, $\frac{w}{16}\times\frac{h}{16}$, and $\frac{w}{32}\times\frac{h}{32}$, respectively.

In this study, two feature extraction networks are adopted to construct two versions of TranSalNet models. One version uses ResNet-50 \cite{RESNET} as an encoder, which is a feature extraction network widely used in saliency prediction. This version of the model is referred to as \textit{TranSalNet\_Res}. The CNN body of ResNet-50 is composed of five convolutional blocks that are denoted as conv1 and conv2$\_$x to conv5$\_$x. We extract feature maps from the deeper conv3$\_$x, conv4$\_$x, and conv5$\_$x blocks. However, \cite{EML-NET} suggests that ResNet-50 itself as an encoder is probably relatively ``shallow." Therefore, we use DenseNet-161~\cite{DenseNet}, which has higher performance on the ImageNet benchmark, as the CNN encoder to build another version referred to as \textit{TranSalNet\_Dense}. For DenseNet-161, it mainly consists of four ``Dense Blocks"  denoted as DenseBlock 1 to 4. We extract feature maps from the deeper DenseBlock 2, DenseBlock 3, and DenseBlock 4.

Although previous work \cite{Huang_2015, ML-Net, DVA_Wang, MSI-Net, GAZEGAN} showed that adopting multi-scale feature maps is beneficial to saliency prediction, our experiments found that using feature maps from shallower network blocks, i.e.\ the conv1 and conv2$\_$x, may cause undesired artefacts to appear in the saliency maps. Therefore, we exclude feature maps from the shallower network blocks.

\subsection{The transformer encoder}

The three sets of multi-scale feature maps are respectively fed into three transformer encoders to enhance the long-range and contextual information. The details of transformer are depicted at the bottom of Figure~\ref{fig:arch}. 
Let $\textbf{x}_{1}$, $\textbf{x}_{2}$, and $\textbf{x}_{3}$ be the feature maps that have spatial dimensions of $\frac{w}{32}\times\frac{h}{32}$, $\frac{w}{16}\times\frac{h}{16}$, and $\frac{w}{8}\times\frac{h}{8}$, respectively, 
first, a $1\times1$ convolution layer ($\text{Conv}_{1\times1}$) is used to reduce the computational cost and align with the acceptable input size of the transformer encoder. 
More specifically, both $\textbf{x}_{1}$ and $\textbf{x}_{2}$ are reduced to 768 dimensions, and $\textbf{x}_{3}$ changed to 512 dimensions. 
\textcolor{black}{Following this, as there is no relative or absolute position information in the feature maps, it is necessary to utilise position embedding (POS) to enable position-awareness before feeding the input into the transformer encoders. Therefore, the absolute POS~\cite{ViT} is implemented before feeding input into the transformer encoders, which performs an element-wise addition to the input and a learnable matrix with the same shape as the input.}
Each transformer encoder contains two same layers of 
standard Multi-head Self-Attention (MSA) and Multi-layer Perceptron (MLP) blocks~\cite{ViT}. 
In our model, we apply 12-heads attention in transformer encoder 1 and 2, and 8-heads in encoder 3.
The MLP block contains two layers with a GELU activation function. Besides, Layer Normalization (LN) and residual connection are applied before and after each block respectively. 
The processing in each transformer encoder can be represented as:

\begin{equation}
\textbf{z}_{0} =\text{Conv}_{1\times1}(\textbf{x}_{i}) \oplus \text{POS}_{i},\; i = 1,2,3
\end{equation}
\begin{equation}
\textbf{z}^{'}_{l} = \text{MSA}(\text{LN}(\textbf{z}_{l-1})) \oplus \textbf{z}_{l-1},\; l = 1,2
\end{equation}
\begin{equation}
\textbf{z}_{l} = \text{MLP}(\text{LN}(\textbf{z}^{'}_{l})) \oplus \textbf{z}^{'}_{l}, \; l = 1,2
\end{equation}
where $\textbf{z}_{l}$ is the output feature maps of the $l$-th layer in transformer encoder, and $\textbf{x}_i$ is the input feature maps from the CNN encoder. 
The feature maps that are passed through transformer encoder 1, 2, and 3 are context-enhanced and denoted as $\textbf{x}^{c}_{1}$, $\textbf{x}^{c}_{2}$, and $\textbf{x}^{c}_{3}$ respectively.

\subsection{The CNN decoder}
A CNN decoder is used to fuse the long-range context-enhanced feature maps from the transformer encoders and restore the original image resolution. The CNN decoder is a fully CNN network containing block$\_$1 to block$\_$7, which is used to implement pixel-level classification to predict saliency maps. 
Batch normalization (BN) and the activation function (ReLU for block$\_$1 to block$\_$6; Sigmoid for block$\_$7) are applied after each $3 \times 3$ convolution operation ($\text{Conv}_{3\times3}$), where the former is used to promote the convergence and the latter is used to increase the nonlinear factor of the model. 
Since the input image is 32-scale downsampled by the encoder network, a 2-scale upsampling that adopts nearest-neighbor interpolation is performed to the feature map in block$\_$1 to block$\_$5 to obtain a saliency map of the same size as the input. In order to enhance the long-range and multi-scale context of the feature map during the decoding process, the upsampled feature map and the transformer's output from the corresponding skip-connection are fused by an element-wise product operation. The processes from block$\_$1 to block$\_$6 can be expressed as:
\begin{numcases}{\textbf{x}^f_i= }
\textbf{x}^{c}_{i}, & $i = 1 $  \\
\text{ReLU}(\text{Upsample}(\hat{\textbf{x}}^f_{i-1}) \odot \textbf{x}^c_{i} ), & $i = 2,3 $\\
\text{Upsample}(\hat{\textbf{x}}^f_{i-1}), & $i = 4,5,6 $
\end{numcases}
\begin{equation}
\hat{\textbf{x}}^f_i = \text{ReLU}(\text{BN}(\text{Conv}_{3\times3}(\textbf{x}^f_i))), \; i =1,2,\dots,6
\end{equation}
where $\textbf{x}^f_i$ and $\hat{\textbf{x}}^f_i$ are the input and output features of the $i$-th block. The output block, i.e., block$\_$7, is used to reduce the dimensionality of the feature maps to a 2D map for pixel-level classification. Therefore, the sigmoid activation function is applied to the feature map:
\begin{equation}
\hat{\textbf{y}} = \text{sigmoid}(\text{Conv}_{3\times3}(\hat{\textbf{x}}^f_6)),
\end{equation}
where $\hat{\textbf{y}}$ is the predicted saliency map.

\subsection{Loss function}

Recent saliency prediction studies \cite{SAM_Cornia,EML-NET,GAZEGAN}  have shown that taking advantage of the saliency evaluation metrics to define the loss function can significantly improve the performance of saliency prediction models. 

Following a similar idea, we adopt a linear combination of four metrics as the loss function to train our model, including the Normalized Scanpath Saliency (NSS), Kullback-Leibler divergence (KLD), Linear Correlation Coefficient (CC), and Similarity (SIM). 
Let $\textbf{y}^s$, $\textbf{y}^f$, and $\hat{\textbf{y}}$ be the ground truth saliency map, fixation map, and predicted saliency map, and $i$ indicates the $i$th pixel of $\textbf{y}^s$ and $\hat{\textbf{y}}$, our loss function is defined as:
\begin{equation}
\begin{split}
L(\textbf{y}^s,\textbf{y}^f,\hat{\textbf{y}}) = & {{\lambda}_1}
L_{\text{NSS}}(\textbf{y}^f,\hat{\textbf{y}})+{{\lambda}_2}L_{\text{KLD}}(\textbf{y}^s,\hat{\textbf{y}})\\&+ {{\lambda}_3}L_{\text{CC}}(\textbf{y}^s,\hat{\textbf{y}})+{{\lambda}_4}L_{\text{SIM}}(\textbf{y}^s,\hat{\textbf{y}}),
\end{split}
\end{equation}
where ${\lambda}_1$, ${\lambda}_2$, ${\lambda}_3$, and ${\lambda}_4$ are the weights of each metric, and
\begin{equation}
L_{\text{NSS}}(\textbf{y}^f,\hat{\textbf{y}}) = \frac{1}{\sum_{i} {\textbf{y}_i^f} }  \sum_{i}\frac{\hat{\textbf{y}}_i - \mu(\hat{\textbf{y}})}{\sigma(\hat{\textbf{y}})}  {\textbf{y}_i^f},
\end{equation}
where $\sigma(\cdot)$ and $\mu(\cdot)$ stand for standard deviation and mean respectively;
\begin{equation}
L_{\text{KLD}}(\textbf{y}^s,\hat{\textbf{y}}) = \sum_{i}\textbf{y}^s_i \log(\epsilon +  \frac{\textbf{y}^s_i}{\epsilon + \hat{\textbf{y}}_i}),
\end{equation}
where $\epsilon$ is a regularization constant and set to $2.2204\times10^{-16}$;
\begin{equation}
L_{\text{CC}}(\textbf{y}^s,\hat{\textbf{y}}) = \frac{\mathrm{cov}(\textbf{y}^s,\hat{\textbf{y}})}{\sigma(\textbf{y}^s) \sigma(\hat{\textbf{y}})},
\end{equation}
where $\mathrm{cov}(\cdot)$ is the covariance and $\sigma(\cdot)$ is standard deviation;
\begin{equation}
L_{\text{SIM}}(\textbf{y}^s,\hat{\textbf{y}}) = \sum_{i} \min(\textbf{y}^s_i,\hat{\textbf{y}}_i).
\end{equation}

In $L_{\text{KLD}}$, $L_{\text{CC}}$ and $L_{\text{SIM}}$, $\textbf{y}^s$, and $\hat{\textbf{y}}$ are normalized so that $\sum_{i} \textbf{y}^s_i$ = $\sum_{i} \hat{\textbf{y}}_i$ = 1.

\textcolor{black}{Since the higher NSS, SIM, and CC values and the lower KLD value represent the better agreement between predicted saliency maps and ground truth, we set ${{\lambda}_1}$, ${{\lambda}_3}$, and ${{\lambda}_4}$ to negative and ${{\lambda}_2}$ to positive. In order to balance the impact of different sub-loss functions on the module result, we determine the weights of individual sub-loss functions based on TranSalNet's performance on the SALICON validation set. In our experiments, the weights are adjusted to ensure these sub-loss functions (note the ranges of output values are different for these functions) contribute relatively equally to the model outcome. This is achieved by training and validating TranSalNet on the SALICON training and validation sets each time by a single sub-loss function. According to the recorded minimal loss values on the validation set, weights are initially assigned to the sub-loss functions so that their contributions to the combined loss are relatively equal. 
In a second step, these weights in a combined loss are further adjusted to achieve balanced results on all evaluation metrics.
As per our empirical studies, the default weights ${{\lambda}_1}$, ${{\lambda}_2}$, ${{\lambda}_3}$, and ${{\lambda}_4}$ of the combined
loss function are set to $-1$, 10, $-2$, and $-1$, respectively.
}

\section{Experimental results and discussion}

\subsection{Datasets}
Four commonly used benchmark saliency datasets are used to train and evaluate our proposed saliency model and variants. 

\begin{itemize}
\item \textit{SALICON}~\cite{salicon2015} contains 10,000 training, 5,000 validation, and 5000 testing images. 
The ground truth annotations of its test set are unpublished and used for a challenge named \textit{LSUN 2017}\footnote{\url{https://competitions.codalab.org/competitions/17136}} to test the performance of the saliency models. 
Different from other benchmark datasets, it employs mouse clicks instead of an eye tracker to record human visual attention.  
\item \textit{CAT2000}~\cite{cat2000} contains 2,000 publicly available images of 20 categories such as action, art, cartoon etc, where each category includes 100 images. Each image is associated with its eye-tracking data of 24 observers. 
\item \textit{MIT1003}~\cite{MIT1003} consists of 1,003 natural indoor and outdoor images with eye-tracking data of 15 observers.
\item \textit{MIT300}~\cite{MIT300} includes 300 natural indoor and outdoor images. The eye-tracking data is unpublished, it is used as the test set of the MIT/Tübingen benchmark~\cite{benchmark}. 

\end{itemize}

\subsection{Evaluation Metrics}
Various metrics have been proposed to evaluate the agreement between the predicted saliency map and the ground truth. In general, these metrics can be described as location-based and distribution-based metrics depending on how the ground truth is represented~\cite{Bylinskii_2019}; the former adopts the fixation map (i.e., in the form of a binary image) and the latter uses the saliency map (i.e., in the form of a gray-scale image) as the ground truth for visual saliency evaluation. Six popular metrics are widely used to quantify the general performance of saliency models, including CC, SIM, KLD, NSS, AUC (Area under ROC Curve), and sAUC (Shuffled AUC). Details of these metrics can be found in ~\cite{Bylinskii_2019}. The first three are distribution-based metrics, and the remaining three are location-based metrics. For KLD, the closer the value is to zero, the better the agreement between prediction and ground truth. For the other five metrics, higher values represent higher consistency.

Now, in this paper, we aims to evaluate the general performance of our proposed model, but in the meantime the perceptual relevance of the saliency model is the focus of our study. To this end, on the basis of the study of~\cite{Bylinskii_2019}, we classify the six metrics into two categories based on their capability of being in close agreement with human judgements of saliency maps: ``\textit{perception-based metrics}", which include NSS, CC, and SIM; and ``\textit{non-perception-based metrics}", which include sAUC, AUC, and KLD~\cite{Bylinskii_2019}. Note, ``\textit{non-perception-based metrics}" do not necessarily mean they are not measuring the gaze behaviour, they may focus on specific properties of viewing behaviour, such as detecting salient objects in the visual field. It is stated in ~\cite{Bylinskii_2019} that ``\textit{AUC, KL are appropriate for detection applications, as they penalize target detection failures. However, where it is important to evaluate the relative importance of different image regions, such as for image-retargeting, compression, and progressive transmission, metrics like NSS or SIM are a better fit.}" This provides sufficient grounds for building perceptually more relevant saliency prediction models, which is the primary goal of our work.

\subsection{Setup}

\label{pra:setup}
By following a similar procedure in the state-of-the-art \cite{DeepGaze2, Huang_2015, DVA_Wang, MSI-Net, SAM_Cornia}, a model should be first initialised by the weights pre-trained on ImageNet \cite{imagenet}, then trained on the 10,000 images of the SALICON training set to reduce the risk of overfitting. Consequently, the best model on its validation set should be selected for further testing on the SALICON test set and training on MIT1003 and CAT2000. 

\begin{table*}[]
%\color{blue}
\caption{\textcolor{black}{Model variants purposely constructed for ablation study to explore the contribution of skip-connections (SC),  transformer encoders ($\text{E}_1$ to $\text{E}_3$), and combined loss function ($L_{\text{CB}}$).}}
\label{tb:variants}
\centering

\begin{tabular}{lcccccc}
\hline
& \multicolumn{6}{l}{} \\ \cline{2-7} 
Model variants &  $\text{E}_1$ & $\text{E}_2$  & $\text{E}_3$  & SC  & Loss function  & Backbone \\ \hline
BaseNet & --  &  -- &  -- &  -- & $L_{\text{BCE}}$  & ResNet-50  \\

BaseNet+ & $\checkmark$  & --  &  -- & --  & $L_{\text{BCE}}$  & ResNet-50 \\

SkipNet & --  & --  & --  &  $\checkmark$ & $L_{\text{BCE}}$  & ResNet-50 \\

TranSalNet\_Res\_BCE & $\checkmark$  & $\checkmark$  & $\checkmark$  & $\checkmark$  & $L_{\text{BCE}}$  & ResNet-50 \\

BaseNet$(L_{\text{CB}})$ & --  &  -- &  -- & --  & $L_{\text{CB}}$  & ResNet-50 \\
BaseNet+$(L_{\text{CB}})$ & $\checkmark$  &  -- & --  & --  & $L_{\text{CB}}$  & ResNet-50 \\
SkipNet$(L_{\text{CB}})$ &--   &--   &--   &  $\checkmark$ &  $L_{\text{CB}}$ & ResNet-50 \\

TranSalNet\_Res &  $\checkmark$ & $\checkmark$  &  $\checkmark$ & $\checkmark$ &  $L_{\text{CB}}$ & ResNet-50 \\
TranSalNet\_Dense & $\checkmark$  &  $\checkmark$ & $\checkmark$  &  $\checkmark$ &  $L_{\text{CB}}$ & DenseNet-161 \\ \hline
\end{tabular}

\end{table*}

To obtain fair results in each dataset, $k$-fold cross-validation ($k=10$) is applied for each model. More specifically, each dataset is divided into 10 non-overlapping subsets. For MIT1003, each subset contains around 100 images; For CAT2000, each subset contains 200 images (10 from each category). Each time, one subset is kept as a test set, one as a validation set, and the remaining eight subsets altogether are used as the training set. To eliminate randomness, each test set corresponds to a fixed validation set and training set. We report the overall performance of 10 times test results.

To reduce the computational cost while aligning with the aspect ratio (4:3) of the images in SALICON, all input images are resized and padded to a same size of 384$\times$288 pixels. 
A consistent standard is followed in all training phases. The Adam optimizer \cite{adam} is used to minimize the loss function. 
The learning rate is set to 
$1 \times 10^{-5}$, 
which is then multiplied by 0.1 for every 3 epochs.
Models are trained with a batch size of 4 for 30 epochs with a stop patience of 5 epochs.

\subsection{Ablation study}

Ablation experiments are conducted to investigate the contribution of three key components in our modelling:
(1) Transformer encoders ($\text{E}_1$, $\text{E}_2$, and $\text{E}_3$ denote Transformer encoder 1, 2, and 3 in Figure~\ref{fig:arch}, respectively),
(2) Skip-connections (SC), 
(3) Combined loss function ($L_{\text{CB}}$). To this end, nine model variants are constructed to demonstrate the added value of one or more of the above components, as shown in Table~\ref{tb:variants}.

\begin{figure*}
\centering
\includegraphics[width=1\textwidth]{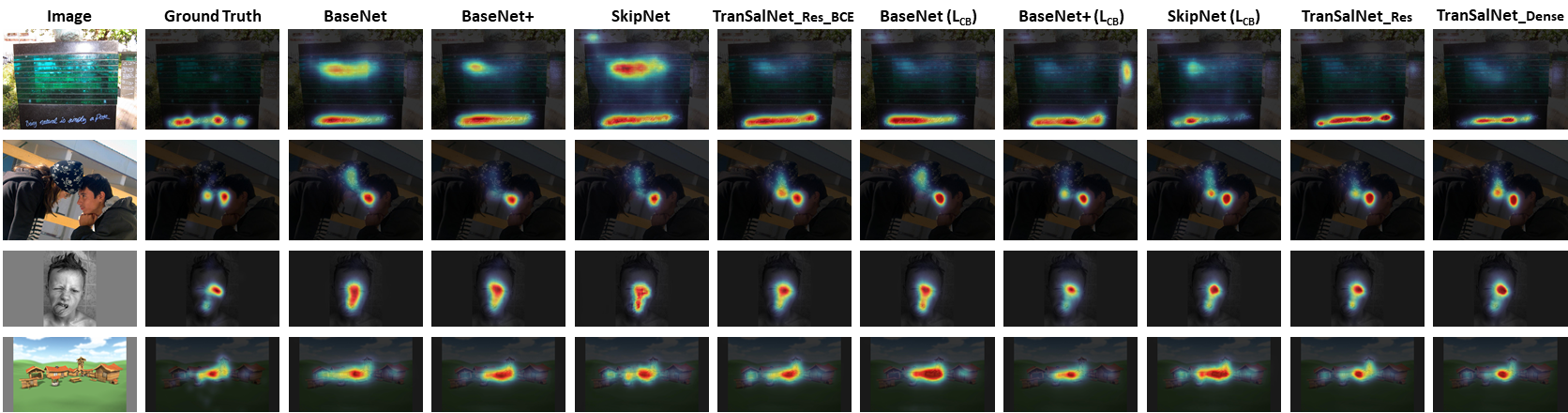}
\caption{Comparison of the saliency prediction performance of nine model variants in our ablation study. The images of top two rows are from the MIT1003 dataset and the bottom two rows are from the CAT2000 dataset. It can be seen that by adopting transformer encoder, skip-connection to provide multi-scale information, and combined loss function, the generated saliency maps are significantly refined relative to the ground truth.}
\label{fig:ab_study}
\end{figure*}

Among them, BaseNet is constructed as a baseline that adopts the widely used ResNet-50 as the CNN encoder, removes all transformer encoders and skip-connections except for the $\text{Conv}_{1\times1}$ layer before transformer encoder 1, and is trained by the BCE loss function. 
BaseNet+ adds the transformer encoder 1 based on the BaseNet.
SkipNet is equipped with skip-connections based on the BaseNet.
TranSalNet\_Res\_BCE adds the transformer encoder 1, 2, and 3 based on the SkipNet, which utilises ResNet-50 as the CNN encoder and is
identical in architecture to the proposed TranSalNet (demonstrated in Figure~\ref{fig:arch}) but is trained by the BCE loss.
The model variants trained by the combined loss that are consistent with the architecture of the above four model variants are denoted as BaseNet$(L_{\text{CB}})$, BaseNet+$(L_{\text{CB}})$, SkipNet$(L_{\text{CB}})$, and TranSalNet\_Res, respectively.
TranSalNet\_Dense replaces ResNet-50 with DenseNet-161 as the CNN encoder.
The overall performance of these model variants on the MIT1003 and CAT2000 datasets is shown in Table~\ref{tb:ab_study}. 
The illustration of saliency maps of four images from these two datasets can also be found in Figure~\ref{fig:ab_study}.

\textcolor{black}{
By comparing BaseNet/BaseNet$(L_{\text{CB}})$ and BaseNet+/BaseNet+$(L_{\text{CB}})$, it can be found that adding a transformer encoder improves the overall performance, i.e., BaseNet+/BaseNet+$(L_{\text{CB}})$ outperforms BaseNet/BaseNet$(L_{\text{CB}})$ in the majority of instances. 
Especially, on the perception-based metrics, i.e., CC, SIM, and NSS, BaseNet+/BaseNet+$(L_{\text{CB}})$ give consistently better performance than BaseNet/BaseNet$(L_{\text{CB}})$, suggesting that the transformer encoder contributes to the perceptual relevance of saliency prediction.
Besides, the benefits of enhancing saliency prediction by providing multi-scale image representations through skip-connections have been demonstrated in previous studies.
Similarly, by adding skip-connections to BaseNet/BaseNet$(L_{\text{CB}})$, the performance of model variants SkipNet/SkipNet$(L_{\text{CB}})$ improves on most instances in the ablation study. 
}

\textcolor{black}{
By uniting transformer encoders and skip-connections, the decoder network can obtain multi-scale feature maps with long-range context enhanced by transformer encoders. As a result, the performance of TranSalNet\_Res\_BCE/TranSalNet\_Res is further boosted on all instances of perception-based metrics as well as most instances of non-perception-based metrics.
This provides additional evidence that the transformer is of added value for visual saliency prediction. Also, this demonstrates the effectiveness of the TranSalNet architecture, which integrates transformer encoders into CNN-based models via skip-connections to obtain multi-scale representations with enhanced long-range visual information.
}
\begin{table*}[]
\caption{Ablation study: performance of nine model variants purposely constructed for ablation study to explore the contribution of skip-connections, transformer encoders, and the combined loss function based on \textbf{MIT1003} and \textbf{CAT2000} datasets. {\color{red}Red} and {\color{orange}orange} font indicate the best and 2nd best performance, respectively.}
\centering
\resizebox{\textwidth}{21mm}{
\begin{tabular}{l|cccccc|cccccc}
\toprule
\multirow{3}{*}{} & \multicolumn{6}{c|}{MIT1003} & \multicolumn{6}{c}{CAT2000}                               \\ \cline{2-13} 
& \multicolumn{3}{c|}{\textbf{perception-based metrics}}            
& \multicolumn{3}{c|}{non-perception-based metrics} 
& \multicolumn{3}{c|}{\textbf{perception-based metrics}}
& \multicolumn{3}{c}{non-perception-based metrics} \\
Variation & CC $\uparrow$ & SIM $\uparrow$ & \multicolumn{1}{c|}{NSS $\uparrow$} & sAUC $\uparrow$ &  AUC $\uparrow$ &  KLD $\downarrow$  & CC $\uparrow$ & SIM $\uparrow$ & \multicolumn{1}{c|}{NSS $\uparrow$} & sAUC $\uparrow$ & AUC $\uparrow$  & KLD $\downarrow$ \\ \midrule\midrule

BaseNet &0.7361&0.5858& \multicolumn{1}{c|}{2.7129} &0.7460&0.9018&0.7886&0.8536&0.7277& \multicolumn{1}{c|}{2.3043} &0.6062&0.8775&0.5190 \\

BaseNet+ &0.7446&0.6003& \multicolumn{1}{c|}{2.7437} &0.7472& 0.9046& 0.8369& 0.8622&0.7343& \multicolumn{1}{c|}{2.3286} &0.6054&0.8784&0.5056 \\

SkipNet  &0.7393& 0.5381& \multicolumn{1}{c|}{2.7571} &0.7505& 0.9011& \textcolor{red}{0.7327}& 0.8582&0.7187& \multicolumn{1}{c|}{2.3298} &\textcolor{red}{0.6111}&0.8782& \textcolor{red}{0.3207} \\

TranSalNet\_Res\_BCE  &0.7593& 0.6122& \multicolumn{1}{c|}{ 2.8326} &0.7507&0.9086& 0.7785&0.8765& 0.7458& \multicolumn{1}{c|}{2.3887} &0.6035& 0.8803& 0.5040 \\

BaseNet$(L_{\text{CB}})$ &0.7386  &0.5966 & \multicolumn{1}{c|}{2.7156 } &0.7508 &0.9046 &0.7839 &0.8589 &0.7332 & \multicolumn{1}{c|}{2.3349 } &0.6063 &0.8787 &0.5146 \\

BaseNet+$(L_{\text{CB}})$ &0.7449&0.6046& \multicolumn{1}{c|}{2.7588 } &0.7505 & 0.9066 & 0.8321 & 0.8666 &0.7381 & \multicolumn{1}{c|}{2.3578 } &0.6076 &0.8794 &0.5053  \\

SkipNet$(L_{\text{CB}})$ &0.7471 & 0.5998 & \multicolumn{1}{c|}{2.7692 } &0.7534 & 0.9067 & \textcolor{orange}{0.7558} & 
0.8612  &0.7347 & \multicolumn{1}{c|}{2.3409} &0.6081 &0.8797 & 0.4730 \\

TranSalNet\_Res &\textcolor{orange}{0.7595}&\textcolor{orange}{0.6145}& \multicolumn{1}{c|}{\textcolor{orange}{2.8501}} &\textcolor{orange}{0.7546}&\textcolor{orange}{0.9093}& 0.7779& \textcolor{orange}{0.8786}& \textcolor{orange}{0.7492}& \multicolumn{1}{c|}{\textcolor{orange}{2.4154}} &0.6054&\textcolor{orange}{0.8811}& 0.5036\\ 

TranSalNet\_Dense &\textcolor{red}{0.7743}&\textcolor{red}{0.6279}& \multicolumn{1}{c|}{\textcolor{red}{2.9214}} &\textcolor{red}{0.7547}&\textcolor{red}{0.9116}& 0.7862& \textcolor{red}{0.8823}& \textcolor{red}{0.7512}& \multicolumn{1}{c|}{\textcolor{red}{2.4290}} &\textcolor{orange}{0.6099}&\textcolor{red}{0.8820}& \textcolor{orange}{0.4715}\\
\bottomrule
\end{tabular}\label{tb:ab_study}
}
\end{table*}

\textcolor{black}{Table~\ref{tb:ab_study} also  demonstrates the practical plausibility of training the proposed model with the linear combination of sub-loss functions.
Compired with the model variations trained by $L_{\text{BCE}}$ (i.e., BaseNet, BaseNet+, SkipNet, and TranSalNet\_Res\_BCE), the model variations trained by $L_{\text{CB}}$ (i.e., BaseNet$(L_{\text{CB}})$, BaseNet+$(L_{\text{CB}})$, SkipNet$(L_{\text{CB}})$, and TranSalNet\_Res) achieve higher performance on the majority of saliency metrics.
In particular, the TranSalNet\_Res outperforms the TranSalNet\_Res\_BCE on all instances in the ablation study. In summary, the effectiveness of the transformer encoder, the TranSalNet architecture, and the combined loss function has now been demonstrated in this ablation study. 
}

In addition, previous research~\cite{EML-NET, DP2E} has shown that using backbones with greater representational capability could improve saliency prediction. Similarly, by simply replacing the backbone from the widely used but comparatively ``shallow" ResNet-50 (used by TranSalNet\_Res) with DenseNet-161~\cite{DenseNet}, TranSalNet\_Dense has been further improved as shown in Table~\ref{tb:ab_study}.

\subsection{Comparison with state-of-the-art methods} 
\subsubsection{On MIT1003 and CAT2000 datasets} 

Seven state-of-the-art deep learning-based saliency models that adopt multi-scale representations or attention mechanisms, including \textcolor{black}{FastSal~\cite{FastSal}, UNISAL~\cite{UNISAL}}, MSI-Net~\cite{MSI-Net}, SAM-VGG~\cite{SAM_Cornia}, SAM-ResNet~\cite{SAM_Cornia}, ML-Net~\cite{ML-Net}, and Deep Visual Attention (DVA)~\cite{DVA_Wang} are selected for the general performance comparison on the MIT1003 and CAT2000 datasets. In order to ensure a fair comparison, the same $k$-fold Cross-Validation ($k=10$ for MIT1003 and CAT2000) strategy and the dataset splitting method used in TranSalNet are employed for fine-tuning and testing of these models. The corresponding pre-trained weights on the SALICON dataset is loaded for each fine-tuning instance. For MIT1003 and CAT2000 datasets, the overall performance of 10 times test results is reported in Table~\ref{tb:perf_comp_mitcat}.

\begin{table*}[]
\caption{Performance comparison of state-of-the-art saliency models on \textbf{MIT1003} and \textbf{CAT2000}. {\color{red}Red} and {\color{orange}orange} font indicate the best and 2nd best performance, respectively.}
\centering
\resizebox{\textwidth}{23mm}{
\begin{tabular}{l|cccccc|cccccc}
\toprule
\multirow{3}{*}{} & \multicolumn{6}{c|}{MIT1003} & \multicolumn{6}{c}{CAT2000}  \\ \cline{2-13} 
& \multicolumn{3}{c|}{\textbf{perception-based metrics}}              
& \multicolumn{3}{c|}{non-perception-based metrics} 
& \multicolumn{3}{c|}{\textbf{perception-based metrics}}
& \multicolumn{3}{c}{non-perception-based metrics} \\
Model name & CC $\uparrow$ & SIM $\uparrow$ & \multicolumn{1}{c|}{NSS $\uparrow$} & sAUC $\uparrow$ &  AUC $\uparrow$ &  KLD $\downarrow$  & CC $\uparrow$ & SIM $\uparrow$ & \multicolumn{1}{c|}{NSS $\uparrow$} & sAUC $\uparrow$ & AUC $\uparrow$  & KLD $\downarrow$ \\ \midrule\midrule

FastSal~\cite{FastSal} &0.5901&0.4783& \multicolumn{1}{c|}{2.0078} &0.7060&0.8745&1.0359
&0.7213&0.6031& \multicolumn{1}{c|}{1.8590} &{\color{orange}0.6180}&0.8599&0.5523 \\

UNISAL~\cite{UNISAL} &0.7340 & 0.5973& \multicolumn{1}{c|}{2.7593} &0.7326&0.9026&1.0138
&0.8417&0.7207& \multicolumn{1}{c|}{2.2575} &0.5982&0.8758&0.5302 \\

MSI-Net~\cite{MSI-Net} &0.7473&0.6081& \multicolumn{1}{c|}{2.8007} &0.7454&0.9068&0.8155&0.8655&0.7398& \multicolumn{1}{c|}{2.3547} &0.6071&0.8809&{\color{red}0.4280} \\
SAM-VGG~\cite{SAM_Cornia} &0.7260&0.5976& \multicolumn{1}{c|}{2.7520} &0.7256&  0.9003& 1.2195& 0.8680& 0.7391& \multicolumn{1}{c|}{2.4138} &0.5966& 0.8784&0.6383 \\
SAM-ResNet~\cite{SAM_Cornia}  &0.7466&0.6068& \multicolumn{1}{c|}{2.8001} &0.7365&0.9024& 1.2470&0.8706&0.7395& \multicolumn{1}{c|}{2.4108} &0.5932& 0.8778& 0.6702 \\
ML-Net~\cite{ML-Net}  &0.5979&0.4960& \multicolumn{1}{c|}{2.3329} &0.7218&0.8623&1.3496&0.5221& 0.5407& \multicolumn{1}{c|}{1.4485} &{\color{red}0.6212}&0.8104& 1.1101 \\
DVA~\cite{DVA_Wang} & 0.6990& 0.5663& \multicolumn{1}{c|}{2.5740} &0.7258&0.8970&{\color{red}0.7528}& 0.8616& 0.7335& \multicolumn{1}{c|}{2.3447} &0.6014 &0.8783& {\color{orange}0.4492} \\
\midrule
TranSalNet\_Res &{\color{orange}0.7595}&{\color{orange}0.6145}& \multicolumn{1}{c|}{{\color{orange}2.8501}} &{\color{orange}0.7546}&{\color{orange}0.9093}& {\color{orange}0.7779}& {\color{orange}0.8786}& {\color{orange}0.7492}& \multicolumn{1}{c|}{{\color{orange}2.4154}} &0.6054&{\color{orange}0.8811}& 0.5036\\ 
TranSalNet\_Dense & {\color{red}0.7743} & {\color{red}0.6279} & \multicolumn{1}{c|}{{\color{red}2.9214}} & {\color{red}0.7547} & {\color{red}0.9116} & 0.7862& {\color{red}0.8823}& {\color{red}0.7512} & \multicolumn{1}{c|}{{\color{red}2.4290}} &0.6099&{\color{red}0.8820}& 0.4715\\
\bottomrule
\end{tabular}\label{tb:perf_comp_mitcat}}
\end{table*}

\begin{figure*}
\centering
\includegraphics[width=1.0\textwidth]{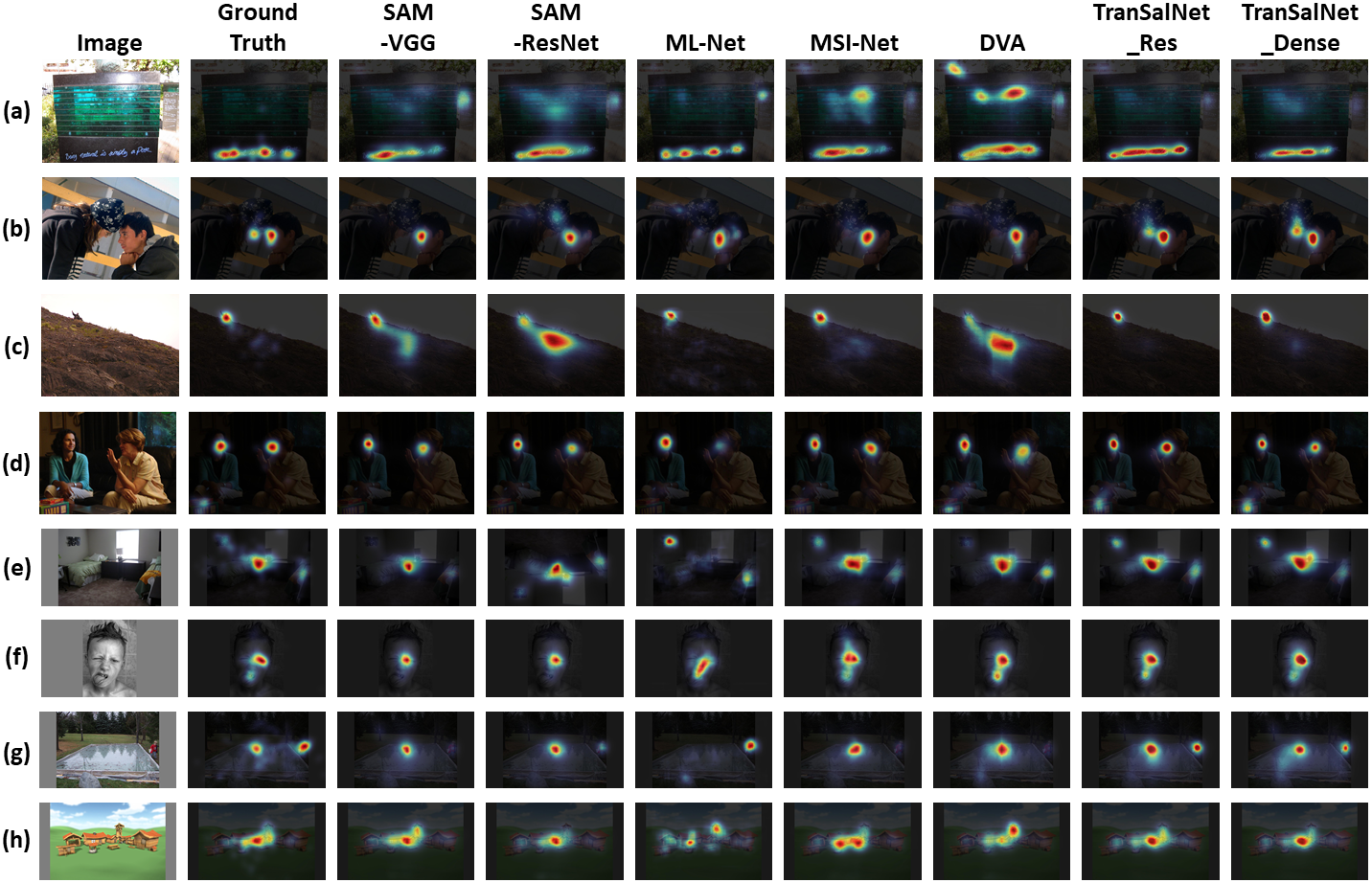}
\caption{Comparison of saliency maps generated by our models (TranSalNet\_Res and TranSalNet\_Dense) and other state-of-the-art saliency models. The images from (a) to (d) are from the MIT1003 dataset, and the images from (e) to (h) are from the CAT2000 dataset.}
\label{fig:demo1}
\end{figure*}

It can be seen that our models (both TranSalNet\_Res and TranSalNet\_Dense) achieve the best performance on all perception-based metrics in both MIT1003 and CAT2000, while producing competitive results on non-perception-based metrics (i.e., being best or 2nd best in most instances in the comparative study). It should be noted that our TranSalNet\_Res and the five state-of-the-art models all use ResNet-50 or VGGNet (representing similar network capacity) as the feature extraction network. TranSalNet\_Res achieves the best performance on most instances (except for sAUC and KLD in CAT2000), implying the contribution of enhanced \textcolor{black}{long-range} information to saliency prediction using transformers. Moreover, the performance our TranSalNet\_Res could be further enhanced by replacing ResNet-50 by a network with higher capacity, namely DenseNet-161.
Figure~\ref{fig:demo1} shows saliency maps generated by our models and other models for images including common contexts such as objects, portraits, natural, indoor, social, and cartoon scenes. By visually assessing these saliency maps, our models are in closer agreement with the ground truth than other models. 

\subsubsection{On MIT300 competition}

For the MIT300 competition, we use the MIT1003 to train an optimal model, in which 703 images are randomly selected as a training set and the rest as a validation set. 
The optimal model is submitted to and tested by the MIT/Tuebingen Saliency Benchmark~\cite{benchmark}. It should be noted that the benchmark evaluates models by different standards, i.e., models must be explicitly claimed as either probabilistic or non-probabilistic models, so they can be fairly evaluated within the category they belong to~\cite{Bylinskii_2019}.
In this paper, same as the original MIT Saliency Benchmark~\cite{Bylinskii_2019}, we ``do not assume that our model is probabilistic". Note that for evaluating probabilistic models, metric-specific adaptations are applied using regularization and scaling of saliency values, hence, a probabilistic model generates optimal saliency maps for individual metrics~\cite{kummerer2018}. But a non-probabilistic model only outputs a single saliency map for all metrics. So it is nontrivial to compare a non-probabilistic (i.e., classical) model to a probabilistic model~\cite{Bylinskii_2019}.
To avoid unfair model comparison under different assumptions, Table~\ref{tb:perf_comp_mit300} shows only non-probabilistic classical saliency models on the leader-board of~\cite{benchmark}. 
It can be seen that our models (both TranSalNet\_Res and TranSalNet\_Dense) consistently rank in the top 1st or 2nd positions on the perception-based metrics (note the only exception is for TranSalNet\_Res on NSS, but its performance score is fairly comparable to the 1st or 2nd scores as shown in Table~\ref{tb:perf_comp_mit300}). On the non-perception-based metrics, our models exhibit competitive performance on sAUC and AUC, with the performance scores comparable to the results in the 1st and 2nd positions. 
In addition, even though we include top probabilistic models such as DeepGaze II-E~\cite{DP2E}, MSI-Net~\cite{MSI-Net}, UNISAL~\cite{UNISAL}, SalFBNet~\cite{SalFBNet}, and DeepGaze II~\cite{DeepGaze2} for performance comparison, our model can still remain competitive in perception-based metrics (results available on website of~\cite{benchmark}).

\subsubsection{On LSUN’17 competition}
Although our aim is to predict the spatial distribution of human fixations, the human attention measured by mouse tracking can still reflect eye movement behaviour to a certain extent~\cite{salicon2015}. SALICON provides so far the largest-scale saliency dataset (via mouse tracking), which allows the opportunity to examine the saliency models from the perspective of being ``data rich". Moreover, for LSUN’17 competition (on SALICON test set), a unified evaluation process is adopted, i.e., the saliency models are not treated differently because of their type of being probabilistic or classical. In the competition each model submitted should generate one single saliency map for each image. Therefore, in order to provide a complementary comparison of state-of-the-art saliency models, Table~\ref{tb:perf_comp_SALICON} reports the results of models submitted to the competition based on the 2017 version (i.e., the latest version). It can be seen that our TranSalNet\_Res and TranSalNet\_Dense achieve superior performance on the perception-based metrics and promising results on other non-perception-based metrics. This shows that our model are competitive on the LSUN 2017 leaderboard, in particular for prediction saliency in a perceptually relevant manner.

\begin{table}[]
\caption{\textbf{MIT300 competition} for saliency models. The results are administered and reported by the Benchmark \cite{benchmark}. {\color{red}Red} and {\color{orange}orange} font indicate the best and 2nd best performance, respectively.}\label{tb:perf_comp_mit300}
\centering
\renewcommand{\arraystretch}{1.2}
\begin{adjustbox}{width=0.48\textwidth}
\begin{tabular}{l|c c c c c c }
\toprule

& \multicolumn{3}{c|}{\textbf{perception-based metrics}}              
& \multicolumn{3}{c}{non-perception-based metrics} \\
Model name & CC $\uparrow$ & SIM $\uparrow$ & \multicolumn{1}{c|}{NSS $\uparrow$} & sAUC $\uparrow$ &  AUC $\uparrow$ &  KLD $\downarrow$ \\ \midrule \midrule 

EML-NET~\cite{EML-NET}& 0.7893 & 0.6756 & \multicolumn{1}{c|}{{\color{red}2.4876}} & 0.7469 & {\color{red}0.8762} & 0.8439 \\
CASNet II~\cite{CASNET2}& 0.7054 &0.5806 & \multicolumn{1}{c|}{1.9859}& 0.7398 & 0.8552 &{\color{red}0.5857} \\
GazeGAN~\cite{GAZEGAN}& 0.7579 & 0.6491 & \multicolumn{1}{c|}{2.2118} & 0.7316 & 0.8607 &1.3390\\
SAM-VGG~\cite{SAM_Cornia}& 0.6630 & 0.5986 &\multicolumn{1}{c|}{1.9552}  &0.7305 &0.8473 &1.2746  \\
SAM-ResNet~\cite{SAM_Cornia}& 0.6897  &0.6122 & \multicolumn{1}{c|}{2.0628}  & 0.7396 & 0.8526 & 1.1710 \\
DVA~\cite{DVA_Wang}& 0.6631 & 0.5848& \multicolumn{1}{c|}{1.9305}& 0.7257 &0.8430 & {\color{orange}0.6293}\\
ML-Net~\cite{ML-Net}&0.6633 & 0.5819 &\multicolumn{1}{c|}{1.9748} & 0.7399 & 0.8386 & 0.8006 \\
eDN~\cite{eDN}&0.4518 &0.4112 &\multicolumn{1}{c|}{1.1399} & 0.6180 &0.8171 & 1.1369 \\
SalGAN~\cite{SalGAN}&0.6740 & 0.5932 &\multicolumn{1}{c|}{1.8620} & 0.7354 & 0.8498 &0.7574\\
HATES&0.7897 & 0.5313 &\multicolumn{1}{c|}{2.3762} & {\color{red}0.7549} & {\color{orange}0.8744} &0.7146\\
\midrule
TranSalNet\_Res & {\color{orange}0.7991} & {\color{orange}0.6852} & \multicolumn{1}{c|}{2.3758} & {\color{orange}0.7471} &0.8730 & 0.9019 \\
TranSalNet\_Dense & {\color{red}0.8070} & {\color{red}0.6895} & \multicolumn{1}{c|}{{\color{orange}2.4134}} & 0.7467 &0.8734 & 1.0141 \\
\bottomrule
\end{tabular}
\end{adjustbox}
\end{table}

\begin{table}[]
\caption{\textbf{LSUN'17 competition (on SALICON test set)} for saliency models. Results are provided by the authors. {\color{red}Red} and {\color{orange}orange} font indicate the best and 2nd best performance, respectively.}\label{tb:perf_comp_SALICON}
\centering
\renewcommand{\arraystretch}{1.2}
\begin{adjustbox}{width=0.48\textwidth}
\begin{tabular}{l|c c c c c c }
\toprule

& \multicolumn{3}{c|}{\textbf{perception-based metrics}}              
& \multicolumn{3}{c}{non-perception-based metrics} \\
Model name & CC $\uparrow$ & SIM $\uparrow$ & \multicolumn{1}{c|}{NSS $\uparrow$} & sAUC $\uparrow$ &  AUC $\uparrow$ &  KLD $\downarrow$ \\ \midrule \midrule

DeepGaze II-E~\cite{DP2E}& 0.872 & 0.733 & \multicolumn{1}{c|}{1.996} & {\color{red}0.767} & {\color{red}0.869} &{\color{orange}0.285}\\
SalFBNet~\cite{SalFBNet}&  0.892 &  0.772 & \multicolumn{1}{c|}{1.952} & 0.740 & {\color{orange}0.868} &{\color{red}0.236}\\
MSI-Net~\cite{MSI-Net}& 0.889 & 0.784 & \multicolumn{1}{c|}{1.931} & 0.736 & 0.865 &0.307\\
UNISAL~\cite{UNISAL}& 0.879 & 0.775 & \multicolumn{1}{c|}{1.952} & 0.739 & 0.864 & -- \\
EML-NET~\cite{EML-NET}& 0.886 & 0.780 & \multicolumn{1}{c|}{{\color{red}2.050}} & 0.746 & 0.866 & 0.520 \\
GazeGAN~\cite{GAZEGAN}& 0.879 & 0.773 & \multicolumn{1}{c|}{1.899} & 0.736 & 0.864 &0.376\\
SAM-ResNet~\cite{SAM_Cornia}& 0.899  &0.793 & \multicolumn{1}{c|}{1.990}  & 0.741 & 0.865 & 0.610 \\
\midrule
TranSalNet\_Res & {\color{orange}0.901} & {\color{orange}0.796} & \multicolumn{1}{c|}{1.998} & 0.742 & 0.866 & 0.414 \\ 
TranSalNet\_Dense & {\color{red}0.907} & {\color{red}0.803} & \multicolumn{1}{c|}{{\color{orange}2.014}} & \textcolor{orange}{0.747} & {\color{orange}0.868} & 0.373 \\ 
\bottomrule
\end{tabular}
\end{adjustbox}
\end{table}

\subsubsection{Discussion}

It is crucial to note that metric selection for saliency model evaluation should be based on specific modelling assumptions and specific target applications~\cite{Bylinskii_2019}. The study in~\cite{Bylinskii_2019} concludes that ``\textit{under the assumptions of non-probabilistic modelling, NSS and CC provide the fairest comparison}"; ``\textit{if evaluating probabilistic models, KLD is recommended}"; and ``\textit{specific tasks and applications also call for a difference choice of metrics}". In~\cite{LIJ}, researchers have verified that ``NSS, CC and SIM best correspond to human perception". In a recent study~\cite{xiaohan}, it is found that CC and SIM are the most appropriate saliency evaluation metrics for image quality assessment applications. Therefore, as the results demonstrated in Table~\ref{tb:perf_comp_mitcat}, Table~\ref{tb:perf_comp_mit300}, and Table~\ref{tb:perf_comp_SALICON}, the proposed saliency models (TranSalNet\_Res and TranSalNet\_Dense) could be the best ``human-like" models (i.e., based on perception-based metrics CC and SIM) to evaluate the relative importance of different image regions for the applications such as image re-targeting, image compression and transmission, and visual quality assessment.

\begin{figure}
\centering
\includegraphics[width=.48\textwidth]{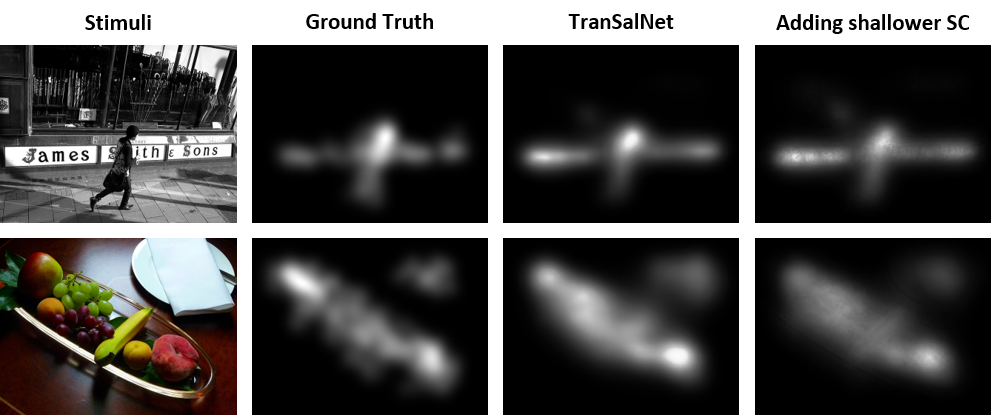}
\caption{\textcolor{black}{The column on the righthand side illustrates the salinency maps with undesired artefacts caused by adding skip-connections to TranSalNet\_Res to connect shallow encoder blocks with decoder block\_4 and block\_5. From left to right, the remaining three columns are: stimuli, ground truth saliency maps, and saliency maps generated from TranSalNet\_Res, respectively. }}
\label{fig:sc}
\end{figure}

\textcolor{black}{Using skip-connections to provide multi-scale features from encoder to decoder has been shown in previous studies to be an effective method for computer vision tasks. For example, the widely used U-Net~\cite{unet} style networks usually connect feature maps of each spatial size to the decoders from shallow to deep encoder blocks. However, as can be seen in Figure~\ref{fig:sc}, using skip-connections to connect shallow encoder blocks (i.e., the blocks provide feature maps with spatial sizes of $\frac{w}{4}\times\frac{h}{4}$ and $\frac{w}{2}\times\frac{h}{2}$) with decoder blocks (i.e., block\_4 and block\_5 in the decoder) may lead to some shapes of objects and texts appearing in the predicted saliency maps, which are not consistent with the ground truth. This implies that adding low-level features from the encoder directly to the decoder may interfere with the saliency prediction of TranSalNet.}

\begin{figure}
\centering
\includegraphics[width=.33\textwidth]{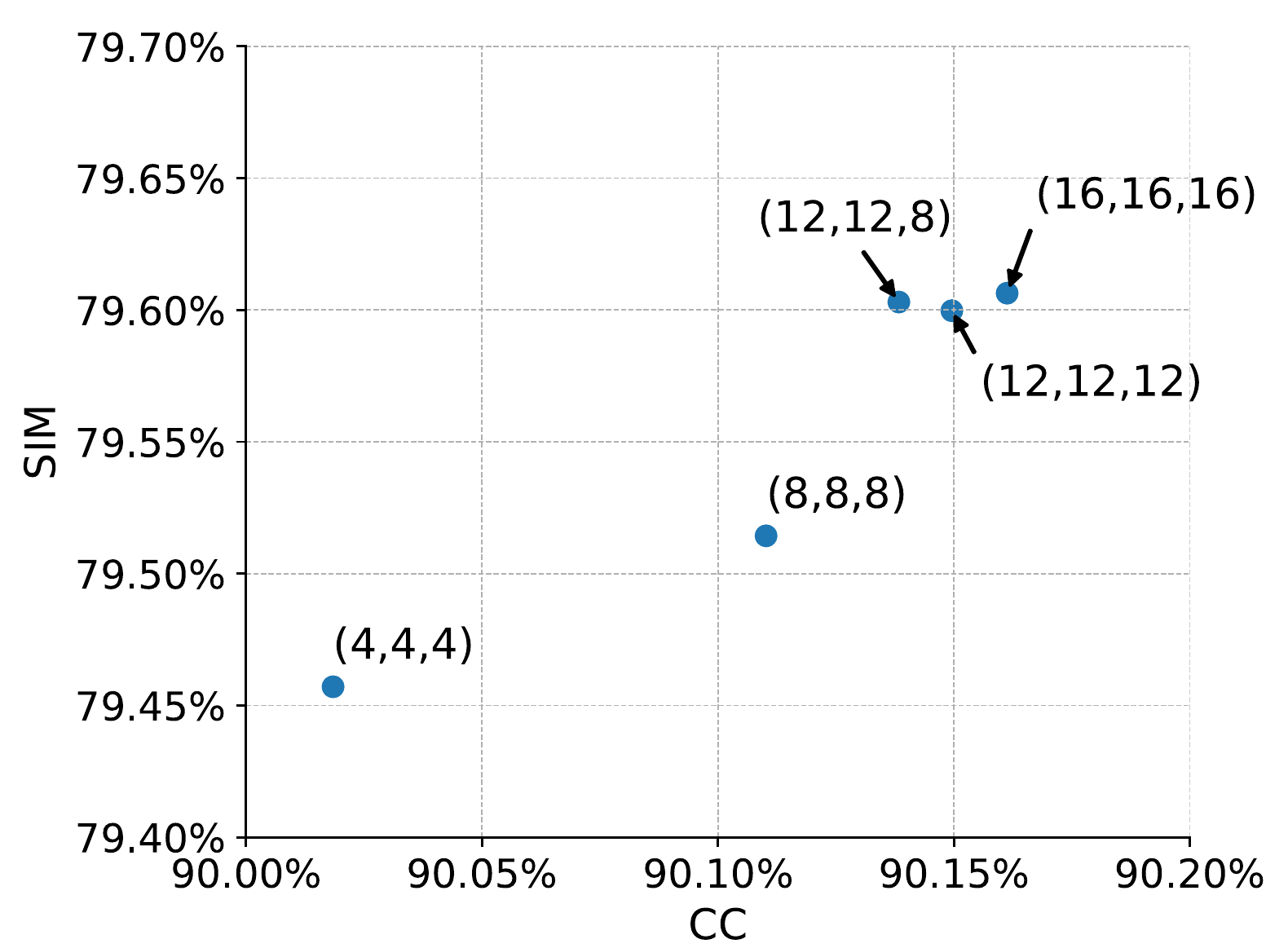}
\caption{\textcolor{black}{Illustrations of the impact of the head number of MSA on the performance of TranSalNet. Values in brackets (i, j, k) indicate the head numbers of transformer encoder 1, 2, and 3 are set to i, j, and k respectively.}}
\label{fig:heads}
\end{figure}

\textcolor{black}{
Multi-head Self Attention (MSA) is part of the transformer encoder. Previous research has shown that the number of heads of MSA could affect the model's performance ~\cite{heads2}. 
According to the suggestions from ~\cite{Bylinskii_2019}, we use CC and SIM as the performance metrics to illustrate the impact of the head number of MSA on our proposed TranSalNet in Figure~\ref{fig:heads}. For each head number combination, the model is trained on the SALICON training set, validated with 2000 images of its validation set, and tested on the rest of the validation set three times. The demonstrated results are the mean results. As can be seen in Figure~\ref{fig:heads}, the scores of CC and SIM tend to increase with the increase in the head number of MSA. However, when the transformer encoders 1 and 2 ($\text{E}_1$ and $\text{E}_2$) adopt 12 heads each, and transformer encoder 3 ($\text{E}_3$) adopts 8 heads of MSA, the performance of the model tends to be saturated in the CC-SIM performance space. Therefore, considering the trade-off between computational resource consumption and model performance, we chose 12 heads for the transformer encoder 1 and 2, and 8 heads for the transformer encoder 3 in this study.
}

\section{Conclusion}

In this paper, we have proposed a novel saliency model for predicting saliency maps that are perceptually in close agreement with the ground truth. By integrating transformers into CNNs, saliency models can significantly benefit from capturing long-range spatial information at multiple perceptual levels. An ablation study has demonstrated the contributions of the transformer encoders to a CNN model, especially the added value of transformers in enhancing the perceptual relevance of saliency prediction. Experimental results show that the proposed models have achieved superior performance on the public benchmarks and competitions for saliency models, particularly having yielded notable results on perception-based saliency evaluation metrics. The perceptually more relevant saliency models have the potential to advance many image processing applications.

\section{Acknowledgments}

This work is funded in part by the Deutsche Forschungsgemeinschaft (DFG, German Research Foundation) – Project-ID 251654672 – TRR 161 and the China Scholarship Council – ID 202008220129.


\end{document}